\documentclass[11pt, draftclsnofoot,onecolumn]{IEEEtran}
\usepackage{bbm}
\usepackage{cite}
\usepackage{amsfonts}
\usepackage{mathrsfs}
\usepackage{graphicx}
\usepackage{amssymb}
\usepackage{latexsym}
\usepackage{amsmath}
\usepackage{stfloats}
\usepackage{cases}
\usepackage{setspace}
\usepackage{bigstrut}
\usepackage{bm}
\usepackage{url}
\usepackage{color}
\usepackage{algorithm}
\usepackage{algorithmic}

\begin{document}

\title{Channel Estimation for Extremely Large-Scale Massive MIMO Systems}
\author{Yu Han, Shi Jin, Chao-Kai Wen, and Xiaoli Ma
\thanks{Y. Han and S. Jin are with the National Mobile Communications Research Laboratory, Southeast University, Nanjing 210096, China (e-mail: hanyu@seu.edu.cn, jinshi@seu.edu.cn.)}
\thanks{C.-K. Wen is with the Institute of Communications Engineering, National Sun Yat-sen University, Kaohsiung 80424, Taiwan (e-mail: chaokai.wen@mail.nsysu.edu.tw).}
\thanks{X. Ma is with the School of Electrical and Computer Engineering, Georgia Institute of Technology, USA (email: xiaoli@gatech.edu).}}
\maketitle

\begin{abstract}
Extremely large-scale massive multiple-input multiple-output (MIMO) has shown considerable potential in future mobile communications. However, the use of extremely large aperture arrays has led to near-field and spatial non-stationary channel conditions, which result in changes to transceiver design and channel state information that should be acquired. This letter focuses on the channel estimation problem and describes the non-stationary channel through mapping between subarrays and scatterers. We propose subarray-wise and scatterer-wise channel estimation methods to estimate the near-field non-stationary channel from the view of subarray and scatterer, respectively. Numerical results demonstrate that subarray-wise method can derive accurate channel estimation results with low complexity, whereas the scatterer-wise method can accurately position the scatterers and identify almost all the mappings between subarrays and scatterers.
\end{abstract}


\section{Introduction}\label{Sec:Introduction}

Extremely large-scale massive multiple-input multiple-output (MIMO) is a promising research direction of multi-antenna technology, in which a significant number of antennas can be widely spread (e.g., on the walls of dense buildings in the city) \cite{Emil2019}. Given that users are surrounded by base station (BS) antennas, extremely large-scale massive MIMO can produce seamless mobile communication services.
However, the use of extremely large aperture array and the close distance between array and user will result in different channel conditions. One difference is the near-field propagation. If users and scatterers are located inside the Rayleigh distance of the array, then the array will experience spherical wavefronts instead of planar wavefronts \cite{Zhou2015,Yin2017,Calvez2018}. Another difference is spatial non-stationarity \cite{Carvalho2019,Ali2019,Amiri2018,Cheng2019}. The channel measurement results of \cite{Carvalho2019} show that different regions of a large aperture array receive varying levels of powers due to the different propagation paths (scatterers) that they can see, whereas some of them cannot see a path. Visible region is introduced to describe the non-stationarity.

Under near-field non-stationary channel conditions, the transceiver should be redesigned, thereby raising new requirements on the acquisition of channel state information. Theoretical analysis found that the signal-to-interference-and-noise ratio of each user increases if the visible regions of different users do not overlap \cite{Ali2019}. Therefore, low-complexity subarray-based transceivers can be designed with the knowledge of the visible regions \cite{Carvalho2019,Ali2019,Amiri2018}. If the positions of the scatterers are further obtained, then enhanced transceiver designs will be enabled. Traditional channel estimation methods, such as least squares (LS) and linear minimum mean square error \cite{Lu2019}, are unable to satisfy these requirements. Although numerous studies have been conducted on positioning the scatterers in near-field stationary channels \cite{Yin2017,Calvez2018}, and more research has begun to focus on the estimation of visible regions of scatterers \cite{Cheng2019}, only a few studies are available on positioning the scatterers and identifying the visible regions simultaneously.

This letter proposes a subarray-wise and a scatterer-wise channel estimation methods to draw up the near-field non-stationary massive MIMO channel. We model the multipath channel with the last-hop scatterers under spherical wavefront and divide the large aperture array into multiple subarrays. Given the stationarity in each subarray, the subarray-wise method treats one subarray individually, and positions the visible scatterers of each subarray on the basis of a refined orthogonal matching pursuit (OMP) algorithm. Utilizing the array gain integrated by multiple subarrays, the scatterer-wise method simultaneously positions each scatterer and detects its visible region to further enhance the positioning accuracy. Numerical results demonstrate that the subarray-wise method achieves an excellent mean square error (MSE) performance with low complexity, whereas the scatterer-wise method can accurately position the scatterers and figure out the non-stationary channel.

\section{System Model}\label{Sec:SystemModel}

\begin{figure}
  \centering
  \includegraphics[scale=0.7]{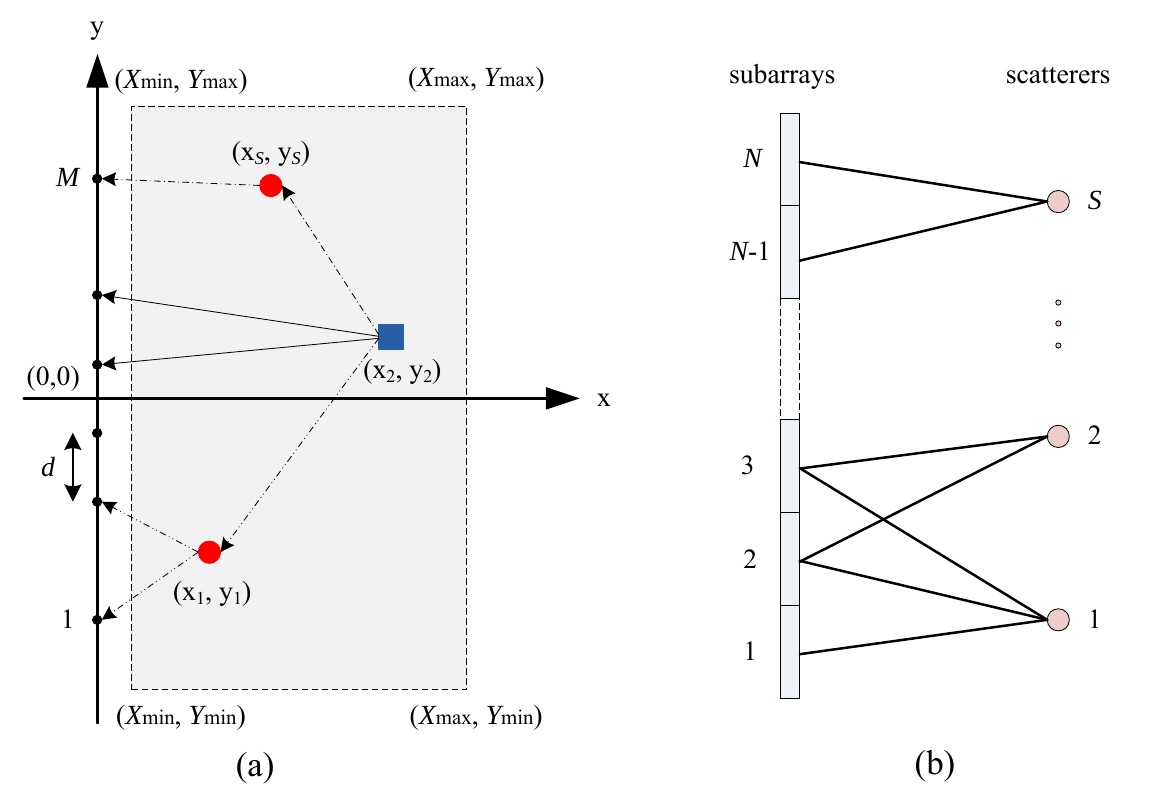}
  \caption{Extremely large-scale massive MIMO. (a) ULA elements (black dots), user (blue square), and scatterers (red circles) in the $xy$ coordinate system. (b) Mapping between subarrays and scatterers (including the user antenna).}\label{Fig:system_model}
\end{figure}

In the extremely large-scale massive MIMO system, the BS is equipped with an $M$-element uniform linear array (ULA)\footnote{ULA is reduced from the more widely used uniform planar array (UPA). Channel estimation methods proposed in this letter can be easily extended to UPA cases, where three-dimensional coordinate system should be considered.}, where $M$ can be $10^3$ or larger. The distance between two adjacent ULA elements is $d$, and the aperture of the ULA is $(M-1)d$.\footnote{Note that in this letter, the distances and the coordinates are normalized by the carrier wavelength.} The ULA center serves as the origin of an $xy$ coordinate system, as shown in Fig.~\ref{Fig:system_model}(a). The ULA lies along the $y$-axis. The coordinate of ULA element $m$ is $(0,(m-1-\frac{M-1}{2})d)$, where $m=1,\ldots,M$. To describe the non-stationary channel, we uniformly divide the ULA into $N$ subarrays, each with $M/N$ antennas.

Users are located in the positive $x$-axis region, and a certain single-antenna user is considered\footnote{Channel estimations of different users are identical and independent from each other. Therefore, we consider the estimation of a single user channel.}. As shown in Fig.~\ref{Fig:system_model}(a), the signal sent by the user may arrive at the ULA along the line-of-sight (LoS) path, or it may be reflected by multiple scatterers. We only focus on the last-jump scatterers of the paths in the visible region of the ULA. User antenna is equivalent to the scatterer for the LoS path. The coordinate of the scatterer $s$ is denoted as $(x_s,y_s)$ to satisfy $X_{\min}<x_s<X_{\max}, Y_{\min}<y_s<Y_{\max}$; $S$ is the number of scatterers (including the user antenna); and $X_{\min}$, $X_{\max}$, $Y_{\min}$, and $Y_{\max}$ draw the bounds of the visible region of the ULA.


\subsubsection{Near-field property}

The wireless signal has spherical wavefront. The array response stimulated by the scatterer at $(x,y)$ is denoted as ${\bf a}(x,y) \in \mathbb{C}^{M\times 1}$, whose $m$th entry is \cite{Zhou2015}
\begin{equation}\label{Eq:Swave}
\left[{\bf a}(x,y)\right]_m = \frac{D_{\rm o}(x,y)}{D_m(x,y)} e^{ j2\pi D_m(x,y) },
\end{equation}
where $D_{\rm o}(x,y) = x$ is the distance between the scatterer and the ULA,
\begin{equation}\label{Eq:D_m}
D_m(x,y) = \sqrt{x^2 + \left(y-d\left(m-(M+1)/2\right)\right)^2}
\end{equation}
is the distance between the scatterer and the ULA element $m$, and $m=1,\ldots,M$.

\subsubsection{Non-stationarity}

When non-stationarity holds, a subarray may not see any scatterer, and a scatterer sees at least one but not all subarrays. Subarray $n$ can receive signals reflected by scatterers (i.e., the visible region of subarray $n$):
\begin{equation}\label{Eq:Psi_n}
\Psi_n = \left\{s_{n,1}, \ldots, s_{n,S_n} \right\},
\end{equation}
which satisfies $1 \le s_{n,i}\le S$ and $0 \le S_n \le S$. Similarly, scatterer $s$ can see subarrays (i.e., the visible region of subarray $n$):
\begin{equation}\label{Eq:Phi_s}
\Phi_s = \left\{n_{s,1}, \ldots, n_{s,N_s} \right\},
\end{equation}
which satisfies  $1 \le n_{s,j}\le N$ and $1 \le N_s \le N$. That is to say, $\Psi_n$ and $\Phi_s$ describe the mapping between subarrays and scatterers from their corresponding views. On the basis of the mapping in Fig.~\ref{Fig:system_model} (b), we can write that $\Psi_1=\{1\}$, $\Psi_2=\Psi_3=\{1,2\}$, $\Psi_{N-1}=\Psi_{N}=\{S\}$, and $\Phi_1=\{1,2,3\}$, $\Phi_2=\{2,3\}$, $\Phi_S=\{N-1,N\}$.


Given the two properties, we model the multipath channel from the user to the BS as
\begin{equation}\label{Eq:ChannelModel}
{\bf h} = \sum_{s=1}^{S} g_s {\bf a}(x_s,y_s) \odot {\bf p}(\Phi_s),
\end{equation}
where $g_s$ is the complex attenuation factor of path $s$, $(x_s,y_s)$ is the coordinate of the $s$th last-jump scatterer, $\odot$ denotes Hadamard product, ${\bf p}_s \in \mathbb{Z}^{M\times 1}$ selects the subarrays that can see the $s$th scatterer with the $m$th entry to be
\begin{equation}\label{Eq:pvec}
\left[{\bf p}(\Phi)\right]_m =
\begin{cases}
1, & \text{if $\lceil \frac{mN}{M}\rceil \in \Phi$,} \\
0, & \text{else,}
\end{cases}
\end{equation}
and $\lceil \cdot \rceil$ rounds a decimal to its nearest higher integer.

During the channel estimation phase, the user sends all-1 pilots to the BS. The pilot received by the BS is ${\bf r} = \sqrt{P}{\bf h} + {\bf w}$, where $P$ is the transmitted power, and ${\bf w} \in \mathbb{C}^{M\times 1}$ is the additive Gaussian noise with zero means and unit variance. Given ${\bf r}$, the BS then estimates channel ${\bf h}$, and more importantly, obtains $(x_s,y_s)$ and $\Psi_n$ or $\Phi_s$ for the subsequent transceiver design or other applications.

\section{Channel Estimation Methods}\label{Sec:Methods}

In this section, we develop two methods to estimate the near-field non-stationary channel. Meanwhile, we find the positions of the scatterers and determine the mapping from the view of subarray and scatterer, respectively.

\subsection{Array pattern}

In traditional far-field stationary massive MIMO systems, when the number of paths is incomparable to the number of antennas, the channel holds sparsity\footnote{The sparsity of the far-field channel is shown in the angular domain instead of the antenna domain ($\bf h$).} and allows us to extract the paths. This sparsity partially comes from the directionality of the array pattern. If the array pattern of the extremely large-scale massive MIMO system still holds directionality, then we can reuse the methods designed for massive MIMO systems to estimate the near-field non-stationary channel.

\begin{figure}
  \centering
  \includegraphics[scale=0.5]{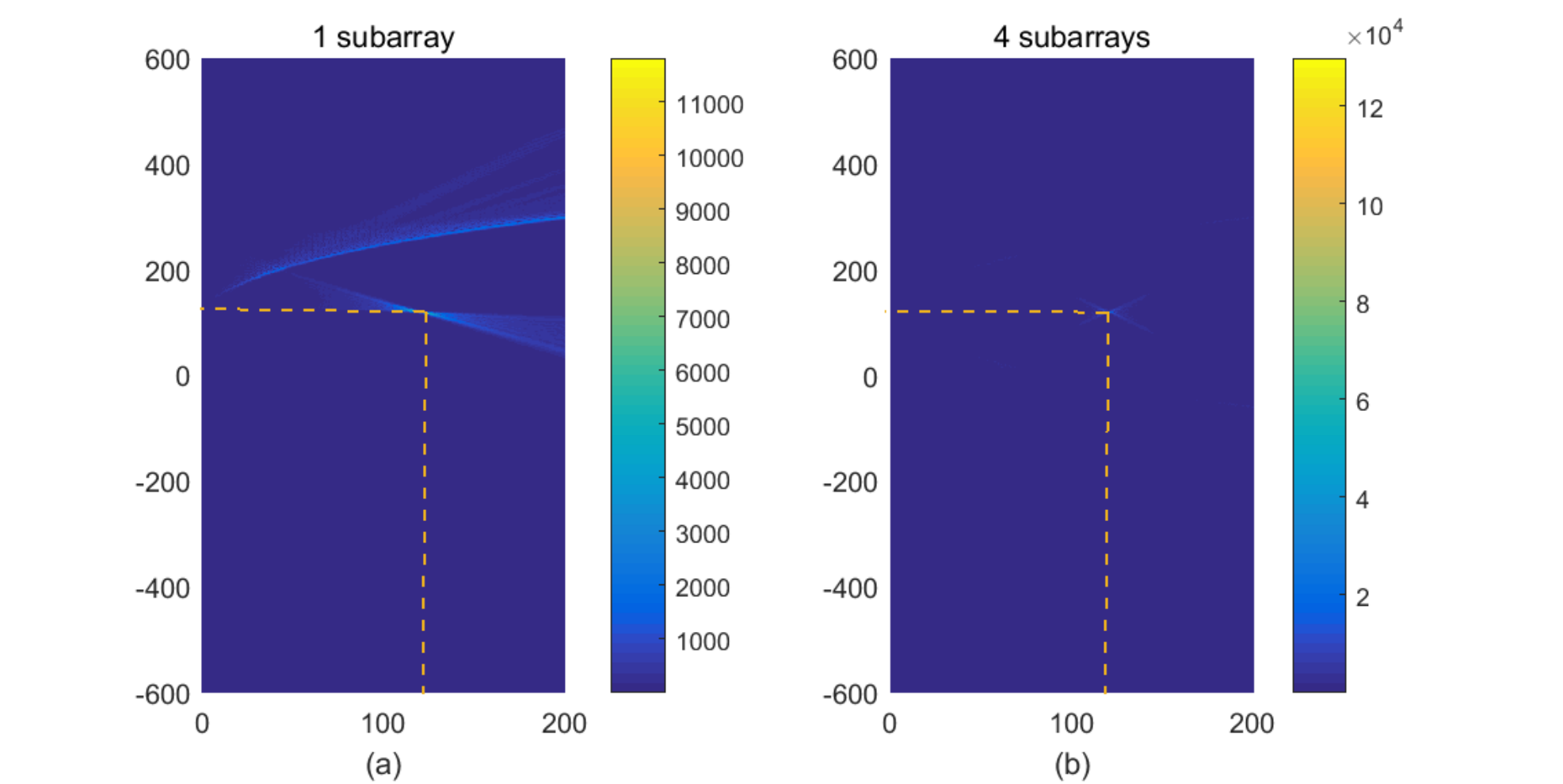}
  \caption{Array pattern of a ULA when (a) 1 subarray and (b) 4 subarrays can see the scatterer at $(120,120)$.}\label{Fig:pattern}
\end{figure}

We define the radiation power from the signal ${\bf z}\in\mathbb{C}^{M\times 1}$ to the $(x,y)$-direction when subarrays in $\Phi$ receive ${\bf z}$ as
\begin{equation}\label{Eq:SwavePattern}
\rho_{\Phi,{\bf z} \to (x,y)} = \left| ({\bf z}\odot {\bf p}(\Phi))^H {\bf b}(x,y)\right|^2,
\end{equation}
where $[{\bf b} (x,y)]_m = e^{ j2\pi D_m(x,y) }$.
The array pattern is obtained by calculating the radiation power from ${\bf z} = {\bf a}(x_s,y_s)$ to all the directions in the visible region of the ULA, i.e.,
\begin{equation}\label{Eq:Fullgrid}
\Xi = \left\{ (x,y)| x = X_{\min}, X_{\min}+ \Delta x,\ldots, X_{\max}; y = Y_{\min}, Y_{\min}+ \Delta y,\ldots, Y_{\max} \right\},
\end{equation}
where $\Delta x$ and $\Delta y$ are the step lengths on the $x$- and $y$-axes, respectively.
Fig.\ref{Fig:pattern} shows the patterns of the 1,024-element ULA, where $N=8$, $d=1/2$, $X_{\min}=0$, $X_{\max}=200$, $Y_{\min}=-600$, $Y_{\max}=600$, $\Delta x=\Delta y=1$, $x_s=y_s=120$, and (a) $\Phi_s=\{4\}$, (b) $\Phi_s=\{2,3,4,5\}$. Despite the different sizes of $\Phi_s$, the strongest radiation power appears exactly at the target position and is considerably larger than that of other positions, demonstrating the directionality of array pattern as well as the sparsity of channel\footnote{The sparsity of the near-field channel is shown in the spatial domain. The spatial domain channel is obtained by projecting the antenna domain channel $\bf h$ to the uniformly sampled $xy$-directions in $\Xi$ as \eqref{Eq:SwavePattern}.} under near-field non-stationary condition. Therefore, sparse signal recovery algorithms such as OMP \cite{Tropp2007} can be applied to extract the paths from the noisy mixture ${\bf r}$.
The channel estimation problem is translated to
\begin{equation}\label{Eq:problem}
\begin{aligned}
\min_{(x_s,y_s),\Phi_s,g_s} &\left\| {\bf r} - \sqrt{P}\sum_{s=1}^{S} g_{s} {\bf a}(x_s, y_s)\odot {\bf p}(\Phi_s) \right\|^2,\\
{\rm s.t.} \quad  &(x_s, y_s)\in\Xi \quad{\rm and} \quad\Phi_s\ne \emptyset.
\end{aligned}
\end{equation}

\subsection{Subarray-wise channel estimation}

Subarray is the smallest unit that experiences stationarity, where the subchannel on subarray $n$ is
\begin{equation}\label{Eq:hn}
{\bf h}_n = \sum_{s\in\Psi_n} g_s {\bf a}_n(x_s,y_s),
\end{equation}
where ${\bf a}_n(x,y)\in\mathbb{C}^{\frac{M}{N}\times 1}$ is the $n$th subvector of ${\bf a}(x,y)$.
If low-complexity subarray-based transceiver designs for traditional stationary systems are adopted, then each subarray is treated individually. Under this consideration, the subarray-wise method is proposed to estimate ${\bf h}_n$ from ${\bf r}_n=\sqrt{P}{\bf h}_n+{\bf w}_n$, where ${\bf r}_n,{\bf w}_n \in\mathbb{C}^{\frac{M}{N}\times 1}$ are the $n$th subvector of ${\bf r}$ and ${\bf w}$, respectively.

OMP is an iterative algorithm that performs in a greedy manner. When applied to channel estimation problem, OMP extracts only one path within each iteration. At the end of the $(s-1)$st iteration, $s-1$ paths are supposedly extracted from ${\bf r}_n$ and the residue is given as follows:
\begin{equation}\label{Eq:Residual1}
{\bf r}^{(s)}_{n,{\rm res}} = {\bf r}_n - \sqrt{P}\sum_{i=1}^{s-1} \tilde g_{n,i} {\bf a}_n(\tilde x_{n,i},\tilde y_{n,i}),
\end{equation}
where $\tilde g_{n,i}$ and $(\tilde x_{n,i},\tilde y_{n,i})$ are the estimates of $g_{s_{n,i}}$ and $(x_{s_{n,i}},y_{s_{n,i}})$, respectively.
In the $s$th iteration, path $s$ is extracted by calculating $\tilde g_{n,s}$ and $(\tilde x_{n,s},\tilde y_{n,s})$ from ${\bf r}^{(s)}_{n,{\rm res}}$ to minimize the residue power.

Coordinate $(\tilde x_{n,s},\tilde y_{n,s})$ is obtained from the grid $\Xi$. In view of the on-grid effect, $\Xi$ should be a dense grid to guarantee the accuracy of $(\tilde x_{n,s},\tilde y_{n,s})$. However, the exhaustive search for a dense grid is time-consuming. Thus, we refine the OMP algorithm by using multilayer grids when extracting each path. We obtain coarse and refined estimates of $(x_{n,s},y_{n,s})$ from a higher- and a lower-layer grid, respectively, and subsequently, estimate $g_s$.

\subsubsection{Coarse estimation}
The higher-layer grid (denoted as $\Xi_{\text H}$) sparsely illustrates the visible region of the ULA. $\Xi_{\text H}$ has the similar expression with \eqref{Eq:Fullgrid}, but larger steps denoted as $\Delta x_{\text H}$ and $\Delta y_{\text H}$, respectively, satisfying $\Delta x_{\text H}\gg \Delta x$, $\Delta y_{\text H}\gg \Delta y$. In the $s$th iteration, we define ${\bf z}_{n,s}\in \mathbb{C}^{M\times 1}$ whose $n$th subvector is ${\bf r}^{(s)}_{n,{\rm res}}$ and zero elsewhere. We select the grid point in $\Xi_{\text H}$, which has the largest radiation power from ${\bf z}_{n,s}$, to be the coarse estimate of the coordinate, that is,
\begin{equation}\label{Eq:HLselect}
(\hat x_{n,s}, \hat y_{n,s}) = \max_{(x,y)\in \Xi_{\text H}} \rho_{\{n\},{\bf z}_{n,s}\to (x,y)}.
\end{equation}

\subsubsection{Refined estimation}
The lower-layer grid $\Xi_{\text L}(\hat x_{n,s},\hat y_{n,s})$ densely illustrates a small region around $(\hat x_{n,s}, \hat y_{n,s})$, where
\begin{multline}\label{Eq:LLgrid}
\Xi_{\text L}(\hat x,\hat y) = \left\{ (x,y)| x = \hat x-\bar X,\ldots,\hat x-\Delta x,\hat x,\hat x+\Delta x,\ldots,\hat x+\bar X; \right. \\ \left. y = \hat y-\bar Y,\ldots,\hat y-\Delta y,\hat y,\hat y+\Delta y,\ldots,\hat y+\bar Y \right\},
\end{multline}
$\bar X$ and $\bar Y$ set the width and height of the low-layer grid. We search the lower-layer grid to obtain the refined estimate of $( x_{n,s}, y_{n,s})$ as
\begin{equation}\label{Eq:LLselect}
(\tilde x_{n,s}, \tilde y_{n,s}) = \max_{(x,y)\in \Xi_{\text L}(\hat x_{n,s},\hat y_{n,s})} \rho_{\{n\}, {\bf z}_{n,s}\to (x,y)}.
\end{equation}
The attenuation factor of path $s$ is estimated by
\begin{equation}\label{Eq:gEst}
\tilde g_{n,s} = g_{\{n\},{\bf z}_{n,s}\to (\tilde x_{n,s}, \tilde y_{n,s})}
\end{equation}
where
\begin{equation}\label{Eq:gFunction}
g_{\Phi,{\bf z} \to (x,y)} = \frac{({\bf a}(x,y)\odot {\bf p}(\Phi))^H {\bf z}}{\left\| {\bf a}(x,y)\odot {\bf p}(\Phi)\right\|^2}.
\end{equation}

\begin{algorithm}[t]
\caption{Subarray-Wise Channel Estimation}\label{alg:separate}
\begin{algorithmic}[1]
\REQUIRE ${\bf r}_1,\ldots,{\bf r}_N$
\ENSURE $\left\{ \tilde{\bf h}_n, (\tilde x_{n,s},\tilde y_{n,s}) \right\},s=1,\ldots,\tilde S_n, n=1,\ldots,N$
\WHILE{$n=1,\ldots,N$}
\STATE Set $s=1$, and ${\bf r}^{(s)}_{n,{\rm res}} = {\bf r}_n$.
\WHILE{\eqref{Eq:OMPstop} does not hold for subarray $n$}
\STATE Coarsely estimate $(\hat x_{n,s},\hat y_{n,s})$ by \eqref{Eq:HLselect}.
\STATE Refine estimate $(\tilde x_{n,s},\tilde y_{n,s})$ by \eqref{Eq:LLselect}.
\STATE Estimate $\tilde g_{n,s}$ by \eqref{Eq:gEst}.
\STATE Set $s=s+1$, and update ${\bf r}^{(s)}_{n,{\rm res}}$.
\ENDWHILE
\STATE Reconstruct $\tilde{\bf h}_n$.
\ENDWHILE
\end{algorithmic}
\end{algorithm}

\subsubsection{Stopping criterion}
The iterations terminate when only the noise remains in the residue. By applying the stopping criterion in \cite{Mamandipoor2016}, the refined OMP algorithm terminates when
\begin{equation}\label{Eq:OMPstop}
\left\| \mathcal{F} \left\{{\bf r}^{(s)}_{n,{\rm res}}\right\} \right\|_\infty \!<\! \log\left(M/N\right) \!-\! \log\log\left( 1\!-\!{(1\!-\!P_{\rm fa})^{-M/N}} \right),
\end{equation}
where $\mathcal{F}\{\cdot\}$ represents the taken Fourier transform and $P_{\rm fa}$ is the false alarm rate. Finally, the BS finds $\tilde S_n$ scatterers that can be seen by subarray $n$. When $\tilde S_n>0$, the channel of subsystem $n$ is reconstructed by
\begin{equation}\label{Eq:tildehn}
\tilde{\bf h}_n = \sum_{s=1}^{\tilde S_n} \tilde g_{n,s} {\bf a}_n(\tilde x_{n,s},\tilde y_{n,s}).
\end{equation}
If $\tilde S_n=0$, then $\tilde{\bf h}_n={\bf 0}$.

The subarray-wise method applies the refined OMP algorithm on ${\bf r}_1,\ldots,{\bf r}_N$ separately, as illustrated in Alg.~\ref{alg:separate}. The estimation results also reflect the mapping between subarrays and scatterers from the view of subarrays. For example, if $(\tilde x_{1,1},\tilde y_{1,1})=(\tilde x_{2,1},\tilde y_{2,1})=(x_1,y_1)$, then scatterer 1 sees subarrays 1 and 2 simultaneously. However, the accuracy of $(\tilde x_{n,s},\tilde y_{n,s})$ is dependent on the scale of a subarray. If $M/N$ is small, then the accuracy will be greatly affected due to the loss of spatial resolution. Consequently, the BS cannot recognize that the two scatterers are the same one. Nevertheless, when the transceiver design on subarray 1 is independent from that on subarray 2, the BS does not need to discover this identity.

\subsection{Scatterer-wise channel estimation}

The joint manipulation of multiple subarrays is a considerably effective method to maximize the array gain and improve the system efficiency. In this condition, the identification of common scatterers in the visible regions of different subarrays becomes essential. The scatterer-wise method estimates the channel directly from the view of scatterers. Fig.~\ref{Fig:pattern} shows that with more subarrays seeing a scatterer, the directionality of the array pattern becomes more distinct. Thus, the scatterer-wise method is designed to position the scatterer by jointly utilizing all the subarrays that can see the scatterer.

The scatterer-wise method is also based on the refined OMP algorithm, as illustrated in Alg.~\ref{alg:joint}. In the $s$th iteration, we estimate the coordinate of scatterer $s$ (i.e., $(x_s, y_s)$) and the attenuation factor $g_s$, and determine the subarrays that can see this scatterer (i.e., $\Phi_s$) from the residue
\begin{equation}\label{Eq:Residual2}
{\bf r}^{(s)}_{\rm res} = {\bf r} - \sqrt{P}\sum_{i=1}^{s-1} \tilde g_{i} {\bf a}(\tilde x_i,\tilde y_i)\odot {\bf p}(\tilde\Phi_i),
\end{equation}
where $\tilde g_i$, $(\tilde x_i,\tilde y_i)$, and $\tilde\Phi_i$ are the estimates of $g_i$, $(x_i,y_i)$, and $\Phi_i$, respectively.
At the beginning of the $s$th iteration (Step 3), the subarrays whose residues do not satisfy \eqref{Eq:OMPstop} are included in $\Phi^{(s)}$. If $\Phi^{(s)}$ is empty, then Alg.~\ref{alg:joint} is terminated. Otherwise, we define ${\bf z}^{(s)} \in\mathbb{C}^{M\times 1}$ whose $n$th subvectors is the $n$th subvector of ${\bf r}^{(s)}_{\rm res}$ if $n\in\Phi^{(s)}$, and zero otherwise.

\subsubsection{Coarse estimation}

We calculate $(\hat x_s,\hat y_s)$, $\hat \Phi_s$, and $\hat g_s$ at Steps 4, 5, and 7. The coordinate is coarsely estimated by
\begin{equation}\label{Eq:HLselect2}
(\hat x_s, \hat y_s) = \max_{(x,y)\in \Xi_{\text H}} \rho_{\Phi^{(s)},{\bf z}^{(s)}\to (x,y)}.
\end{equation}

Then, we derive $\hat\Phi_s$ from $(\hat x_s,\hat y_s)$. $\hat\Phi_s$ is a subset of $\Phi_s$. Subarrays in $\hat\Phi_s$ exhibit the largest radiation power from ${\bf z}^{(s)}$ to $(x_s,y_s)$. In Step 5, we calculate the radiation power at each subarray in $\Phi^{(s)}$ and normalize these power values by
\begin{equation}\label{Eq:gammat}
\gamma_n = \frac{\rho_{\{n\},{\bf z}^{(s)} \to (\hat x_s,\hat y_s)}}{\sum_{j\in\Phi^{(s)}} \rho_{\{j\},{\bf z}^{(s)} \to (\hat x_s,\hat y_s)}}
\end{equation}
where $n\in\Phi^{(s)}$. The sum of the largest $\hat N_s$ values of $\gamma_t$ is assumed to be beyond a threshold $\delta$, where $0<\delta<1$. Then, the corresponding $\hat N_s$ subarrays are included in $\hat\Phi_s$.

We do not estimate $\hat g_s$ on basis of $(\hat x_s,\hat y_s)$ and $\hat \Phi_s$ because the accuracy of $\hat g_s$ will be low. Instead, we estimate $\hat g_s$ after refining $(\tilde x_s,\tilde y_s)$. We define $\hat{\bf z}_s \in\mathbb{C}^{M\times 1}$ whose $n$th subvectors is $n$th subvector of ${\bf r}^{(s)}_{\rm res}$ if $n\in\hat\Phi_s$, and zero otherwise. $\hat g_s$ is calculated by
\begin{equation}\label{Eq:hatg}
\hat g_s = g_{\hat\Phi_s,\hat{\bf z}_s \to (\tilde x_s,\tilde y_s)}.
\end{equation}

\subsubsection{Refined estimation}
We refine the coarse estimates to $(\tilde x_s,\tilde y_s)$, $\tilde \Phi_s$, and $\tilde g_s$ at Steps 6, 8, and 9, respectively. The coordinate of scatterer $s$ is refined by
\begin{equation}\label{Eq:LLselect2}
(\tilde x_s, \tilde y_s) = \max_{(x,y)\in \Xi_{\text L}(\hat x_s,\hat y_s)} \rho_{\hat\Phi_s, \hat{\bf z}_s\to (x,y)}.
\end{equation}

Then, we estimate $\tilde\Phi_s$. Given the directionality of the array pattern, we can derive that for subarray $n\in\Phi^{(s)}$, it holds that
\begin{equation}\label{Eq:eRho}
\mathbb{E}\left[\rho_{\{n\},{\bf z}^{(s)} \to (\tilde x_s,\tilde y_s)}\right] =
\begin{cases}
\mathbb{E}\left[ \left|g_s {\bf b}^H_n(\tilde x_s,\tilde y_s) {\bf a}_n( x_s, y_s) \right|^2 + \left\|{\bf w}_{n}\right\|^2 \right], & \text{if $n \in \Phi_s$,} \\
\mathbb{E}\left[ \left\|{\bf w}_{n}\right\|^2 \right] , & \text{else,}
\end{cases}
\end{equation}
where ${\bf b}_n(x,y)$ is the $n$th subvector of ${\bf b}(x,y)$, $\mathbb{E}\left[\cdot\right]$ means taking expectation, and $\mathbb{E}\left[ \left\|{\bf w}_{n}\right\|^2 \right] = M/N$. By applying $(\tilde x_s,\tilde y_s)$ and $\hat g_s$ into \eqref{Eq:eRho} and considering the estimation errors of these parameters, we define that for subarray $n\in\Phi^{(s)}$, if
\begin{equation}\label{Eq:judgeNS}
\rho_{\{n\},{\bf z}^{(s)} \to (\tilde x_s,\tilde y_s)} \ge \alpha \left| \hat g_s {\bf b}^H_n(\tilde x_s,\tilde y_s) {\bf a}_n( \tilde x_s, \tilde y_s) \right|^2 + M/N,
\end{equation}
is satisfied, where $0<\alpha<1$, then $n$ is included in $\tilde \Phi_s$.

We define $\tilde{\bf z}_s \in\mathbb{C}^{M\times 1}$ whose $n$th subvectors is $n$th subvector of ${\bf r}^{(s)}_{\rm res}$ if $n\in\tilde\Phi_s$, and zero otherwise. Then, $\tilde g_s$ is calculated by
\begin{equation}\label{Eq:tildeg}
\tilde g_s = g_{\tilde\Phi_s,\tilde{\bf z}_s \to (\tilde x_s,\tilde y_s)}.
\end{equation}

When the Alg.~\ref{alg:joint} is terminated, we apply $\tilde g_s$, $(\tilde x_s,\tilde y_s)$, and $\tilde\Phi_s$ into \eqref{Eq:ChannelModel} and obtain the reconstructed channel $\tilde{\bf h}$.

\begin{algorithm}[t]
\caption{Scatterer-Wise Channel Estimation}\label{alg:joint}
\begin{algorithmic}[1]
\REQUIRE ${\bf r}_1,\ldots,{\bf r}_N$
\ENSURE $\tilde {\bf h},\left\{\tilde g_s,(\tilde x_s,\tilde y_s),\tilde \Phi_s\right\},s=1,\ldots,\tilde S$
\STATE Set $s=1$, and ${\bf r}^{(s)}_{\rm res} = {\bf r}$.
\WHILE{\eqref{Eq:OMPstop} does not hold for at least one subarray}
\STATE Include subarrays that do not satisfy \eqref{Eq:OMPstop} into $\Phi^{(s)}$.
\STATE Coarsely estimate $(\hat x_s,\hat y_s)$ by \eqref{Eq:HLselect2}.
\STATE Coarsely estimate $\hat\Phi_s$ by observing \eqref{Eq:gammat}.
\STATE Refine estimate $(\tilde x_s,\tilde y_s)$ by \eqref{Eq:LLselect2}.
\STATE Coarsely estimate $\hat g_s$ by \eqref{Eq:hatg}.
\STATE Refine estimate $\tilde \Phi_s$ by observing \eqref{Eq:judgeNS}.
\STATE Refine estimate $\tilde g_s$ by \eqref{Eq:tildeg}.
\STATE Set $s=s+1$, and update ${\bf r}^{(s)}_{\rm res}$.
\ENDWHILE
\STATE Reconstruct $\tilde{\bf h}$.
\end{algorithmic}
\end{algorithm}

The complexity of each step in the two methods are compared in Table I, where $|\Xi_{\text H}|$, $|\Xi_{\text L}|$, $|\Phi^{(s)}|$, $|\hat\Phi_s|$, and $|\tilde\Phi_s|$ denote the sizes of $\Xi_{\text H}$, $\Xi_{\text L}$, $\Phi^{(s)}$, $\hat\Phi_s$, and $\tilde\Phi_s$, respectively. Given that $\max(\tilde S_n) \le \tilde S$, it satisfies $\sum_{n=1}^N\tilde S_n/N \le \tilde S$. Moreover, the subarray-wise method does not need to estimate $\hat\phi$, $\hat g$, and $\tilde g$. Therefore, the complexity of the subarray-wise method is smaller than that of the scatterer-wise method. The subarray-wise method estimate ${\bf h}_n$ simply from ${\bf r}_n$. Whereas the scatterer-wise method estimate ${\bf h}_n$ from ${\bf r}$, which contains the interference from other subarrays. Thus, the subarray-wise method obtains more accurate channel estimation results. On the other hand, the scatterer-wise method utilizes the multiple subarray gain to estimate the positions and visible regions of the scatters, thereby achieving more accurate positioning and mapping results. Therefore, the subarray-wise method suits for the low-complexity subarray-based transceiver design. The scatterer-wise method enables substantially efficient and comprehensive globalized transceiver design.

\begin{table}\label{tab:complexity}
  \centering
  \caption{Computational Complexity}
  \begin{tabular}{|c|c|c|}
  \hline
  Step & Subarray-wise & Scatterer-wise\\
  \hline
  Stopping criterion & $O(\sum_{n=1}^N {\tilde S_n \frac{M}{N}\log_2\frac{M}{N}})$ & $O({\tilde S M\log_2M})$ \\
  \hline
  Estimate $(\hat x,\hat y)$ & $O(\sum_{n=1}^N {\tilde S_n |\Xi_{\text H}|\frac{M}{N}}) $ & $O({\tilde S |\Xi_{\text H}| M}) $\\
  \hline
  Estimate $\hat\phi$ & 0 & $O({\tilde S} {|\Phi^{(s)}|M})$\\
  \hline
  Estimate $\hat g$ & 0 & $O(\tilde S M)$\\
  \hline
  Estimate $(\tilde x,\tilde y)$ & $O(\sum_{n=1}^N {\tilde S_n |\Xi_{\text L}|\frac{M}{N}}) $ & $O({\tilde S |\Xi_{\text L}| M}) $\\
  \hline
  Estimate $\tilde\phi$ & 0 & $O({\tilde S} {|\hat\Phi_s|M})$\\
  \hline
  Estimate $\tilde g$ & $O(\sum_{n=1}^N {\tilde S_n M})$ & $O(\tilde S M)$\\
  \hline
  \end{tabular}
\end{table}

\section{Numerical Results}\label{Sec:NumericalResults}

In this section, we evaluate the performances of the proposed channel estimation methods. We set $M=1,024$, $d=1/2$, $X_{\min}=20$, $X_{\max}=200$, $Y_{\min}=-600$, and $Y_{\max}=600$. $S=2$ scatterers are randomly distributed in the visual region of the ULA, and $N_s=N/2$ and $0.5<|g_s|^2<1$ holds for $s=1,\ldots,S$. We set $\Delta x_{\rm H}=\Delta y_{\rm H}=4$, $\Delta x_{\rm L}=\Delta y_{\rm L}=0.1$, and $P_{\rm fa}=0.01$ for the hierarchical grids, and $\delta=0.5$, $\alpha=0.8$ for the scatterer-wise method.

\begin{figure}
  \centering
  \includegraphics[scale=0.44]{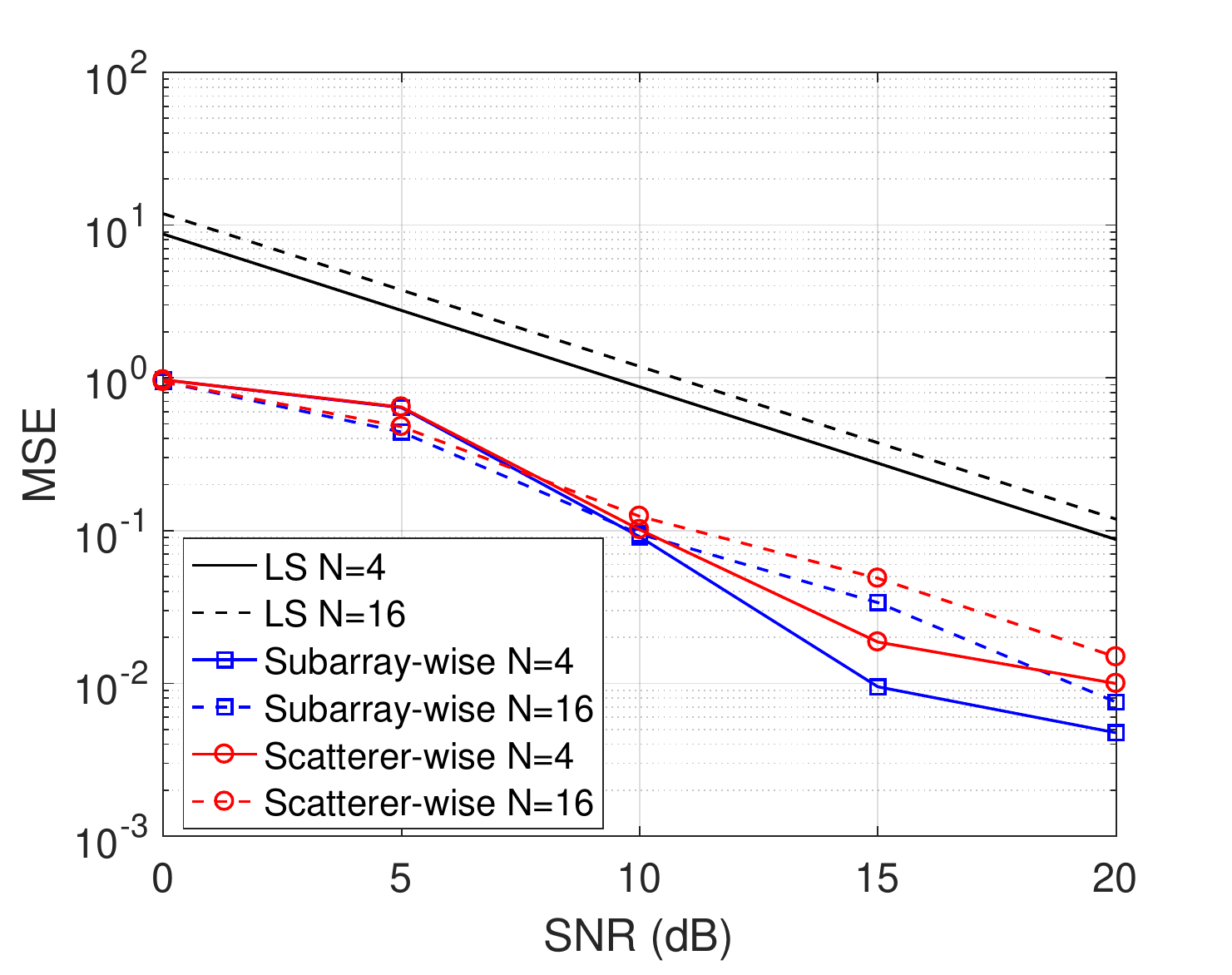}
  \caption{Comparisons of MSE performance between the channel estimation methods and the LS method.}\label{Fig:MSE}
\end{figure}

Fig.~\ref{Fig:MSE} compares the MSE performances of the proposed methods with that of the LS estimation when $N\in\{4, 16\}$. The linear value of signal-to-noise ratio (SNR) equals $P$. We only consider the subarrays where $\Psi_n \ne \emptyset$, and the MSE of the estimated channel $\tilde{\bf h}_n\in\mathbb{C}^{\frac{M}{N}\times 1}$ is calculated as $\mathbb{E}[{\|\tilde{\bf h}_n -{\bf h}_n\|^2}/{\|{\bf h}_n\|^2}]$. Notably, MSE of the LS method is larger than $10^{-2}$ at SNR $= 20$ dB because ${D_0(x,y)}/{D_m(x,y)}\le 1$ holds in \eqref{Eq:Swave}. LS has the drawback of increasing the noise. The proposed methods achieve considerably lower MSEs than the LS method, and can position the scatterers and determine the non-stationarity, thereby showing great advantage than LS. In high SNR region, despite the lower complexity of the subarray-wise method, it exhibits better MSE performance than the scatterer-wise method, which is in accordance with the analysis in Section \ref{Sec:Methods}.

\begin{figure}
  \centering
  \includegraphics[scale=0.44]{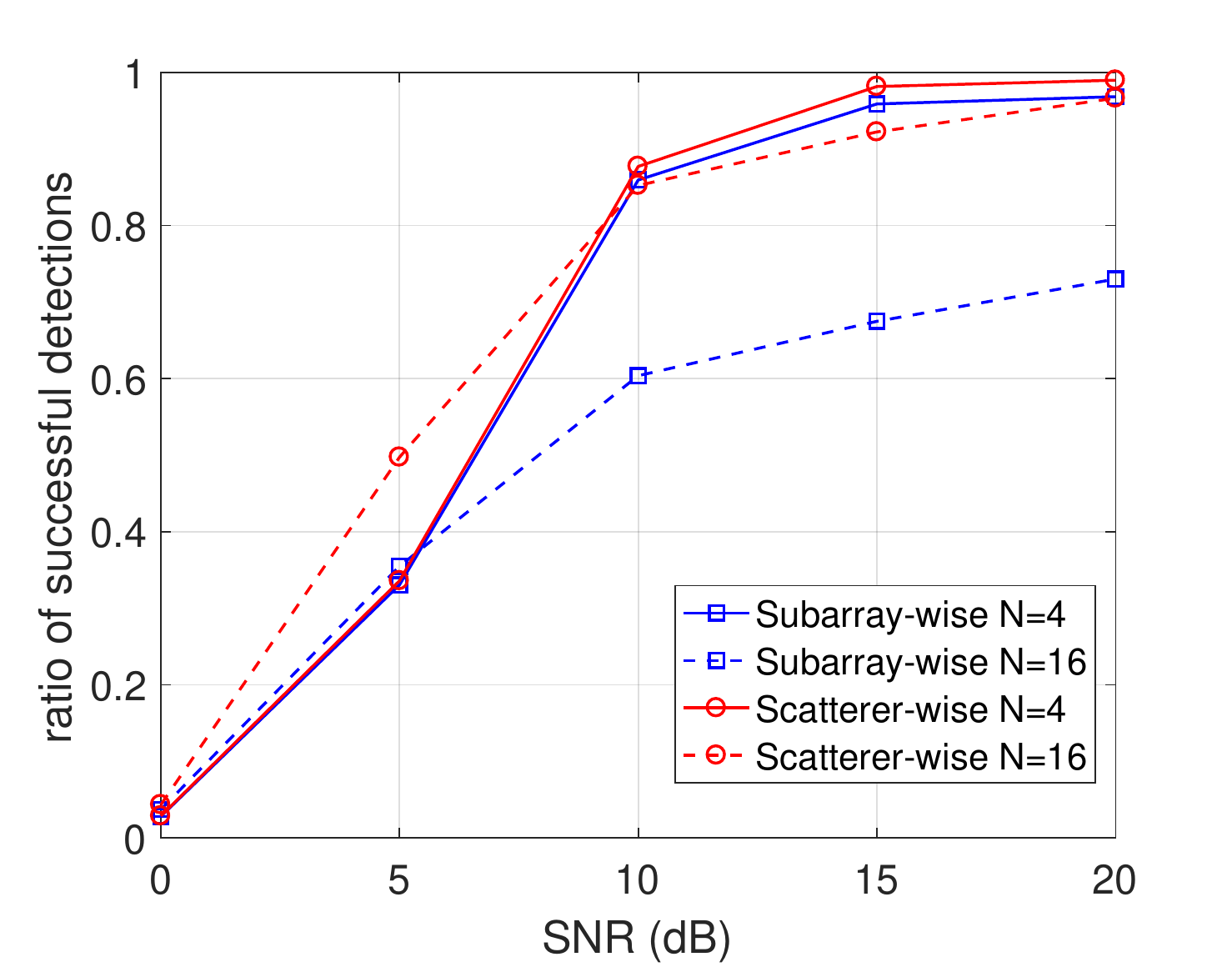}
  \caption{Ratios of successful detections of the two channel estimation methods under different settings of $N$.} \label{Fig:SuccessRatio}
\end{figure}

Fig.~\ref{Fig:SuccessRatio} further evaluates the accuracy of the scatterer positioning and non-stationary mapping results of the two methods under the same condition illustrated in Fig.~\ref{Fig:MSE}. The mapping between subarray $n$ and scatterer $s$ is successfully detected if one detected scatterer $t$ exists, the Euclidean distance of which with $(x_s, y_s)$ is less than 10 (normalized by wavelength), and scatterer $t$ can be seen by subarray $n$.
When SNR is less than 10 dB, both methods cannot work well. With the increase in SNR, the subarray-wise method can accurately position the scatterers when $N=4$, that is, the scale of each subarray is large. Meanwhile, the scatterer-wise method always outperforms the subarray-wise method. Notably, the performance gap between the two methods is enlarged with the increase in $N$. This is because the accuracy of the refined OMP algorithm is degraded due to the reduction of the subarray scale; however, the scatterer-wise method maintains the high positioning and mapping accuracy results with the integrated array gain. When $N=16$ and ${\rm SNR}\ge15$ dB, the successful detection ratio of subarray-wise method exceeds 0.9, which demonstrates the effectiveness of the scatterer-wise method.

\section{Conclusion}\label{Sec:Conclusion}

This letter introduced a channel model that describes the near-field non-stationary properties in the extremely large-scale massive MIMO system. Two channel estimation methods were proposed to position the scatterers and identify the mapping between subarrays and scatterers. The subarray-wise method designed for subarray-based transceivers utilized the stationary on a subarray and positioned the visible scatterers of each subarray. The scatterer-wise method designed for joint subarray transceivers utilized the array gain to position the scatterer and determine its mapping with subarrays simultaneously. The numerical results demonstrated that the low-complexity subarray-wise method exhibits better MSE performance, whereas the scatterer-wise method can accurately position the scatterers and find almost all the mappings.

\end{document}